\documentclass{article}
\usepackage{dcase2016,amsmath,graphicx,url,times,hyperref}
\title{HIERARCHICAL LEARNING FOR DNN-BASED ACOUSTIC SCENE CLASSIFICATION}

\name{Yong Xu, Qiang Huang, Wenwu Wang, Mark D. Plumbley\thanks{This work was supported by the Engineering and Physical Sciences Research Council (EPSRC) of the UK under the grant EP/N014111/1.}}
\address{Centre for Vision, Speech and Signal Processing, University of Surrey, UK\\
	\text{\{yong.xu, q.huang, w.wang, m.plumbley\}@surrey.ac.uk{\scriptsize }}
	}
\begin{document}

\ninept
\maketitle

\begin{sloppy}

\begin{abstract}
In this paper, we present a deep neural network (DNN)-based acoustic scene classification framework. Two hierarchical learning methods are proposed to improve the DNN baseline performance by incorporating the hierarchical taxonomy information of environmental sounds. Firstly, the parameters of the DNN are initialized by the proposed hierarchical pre-training. Multi-level objective function is then adopted to add more constraint on the cross-entropy based loss function. A series of experiments were conducted on the Task1 of the Detection and Classification of Acoustic Scenes and Events (DCASE) 2016 challenge. The final DNN-based system achieved a 22.9\% relative improvement on average scene classification error as compared with the Gaussian Mixture Model (GMM)-based benchmark system across four standard folds.
\end{abstract}


\section{Introduction}
\label{sec:intro}

In recent years, much research effort has been attracted for making sense of everyday or environmental sounds. It focuses on how to convert audio (non-speech and non-music) recordings into understandable and actionable information: specifically how to allow people to search, browse and interact with sounds. Some specific tasks were investigated in recent years, including acoustic scene classification (ASC) \cite{barchiesi2015acoustic}, sound event detection (SED) \cite{mesaros2010acoustic, zhuang2010real} and domestic audio tagging. ASC aims to associate a semantic label to an audio segment that identifies the sound environment where it has been produced \cite{barchiesi2015acoustic}. The goal of SED is to detect the sound events that are present within an audio signal, estimate their start and end times, and give a class label to each of the events. For audio tagging, there is no information about sound event onset or offset, only labels. This paper will focus on the ASC task.

The ASC problem was first proposed by Sawhney and Maes \cite{sawhney1997situational}. Recently, more related work was conducted during the IEEE AASP Challenge: Detection and Classification of Acoustic Scenes and Events \cite{giannoulis2013ieee, giannoulis2013detection, stowell2015detection}. Mel Frequency Cepstrum Coefficients (MFCCs) were used as the audio feature by most of the submitted systems. GMMs, Support Vector Machines (SVMs) or hidden Markov models (HMMs) were commonly used classifier \cite{giannoulis2013detection, geiger2013large, lee2013acoustic}. Other methods, such as non-negative matrix factorization (NMF) approaches can also be used to extract an intermediate representation prior to classification \cite{bisot2016acoustic}.

Recently, deep learning methods have obtained great successes in speech, image and video fields \cite{xu2014experimental, xu2015regression, hinton2012deep, krizhevsky2012imagenet} since Hinton and Salakhutdinov \cite{hinton2006reducing} showed the insights in using a greedy layer-wise unsupervised learning procedure to train a deep model in 2006. Deep learning methods were also investigated for acoustic scene classification tasks in \cite{petetin2015deep, piczak2015environmental, ravanelli2014audio}. In \cite{petetin2015deep}, a series of experimental investigations on the DNN structure, including the number of hidden layers and input frame expansion, were presented. It also demonstrated that DNN can yield better results than GMM and SVM. Convolutional neural networks (CNNs) which are the variant of DNNs have been also adopted for environmental sound classification in \cite{piczak2015environmental}.

There has also been research about the taxonomy of the environmental sounds \cite{niessen2010categories, salamon2014dataset}. The taxonomy of environmental sounds indicates that hierarchical categories information exists in sound classes. For example, environmental sounds can be coarsely classified into \textit{indoor}, \textit{outdoor} and \textit{vehicle} in Fig. \ref{fig:taxonomy}, and these are the high-level scene classes. Meanwhile, corresponding branches denote the low-level scene classes.

In this paper, we propose a hierarchical learning method incorporating the acoustic scene taxonomy information for ASC in a DNN-based framework. Two approaches are presented to utilize the acoustic scene taxonomy information. Firstly, a high-level DNN is discriminatively trained to predict three high-level classes, namely \textit{vehicle}, \textit{indoor} and \textit{outdoor}. Then the trained DNN is used to initialize the low-level DNN except for the top classification layer to learn the more difficult low-level scene classes, namely \textit{bus}, \textit{home}, \textit{park}, etc. This learning process is named as \textbf{hierarchical pre-training}, which follows the common ``easiest thing first hardest second" learning experience of human \cite{lee2011learning}. Hierarchical pre-training is a supervised process which is different from the common Restricted Boltzmann machine (RBM) based unsupervised pre-training \cite{hinton2006reducing}. The second idea is based on a proposed \textbf{multi-level objective function}, which means the DNN not only predicts the target low-level scene classes, but also predicts the three high-level scene classes as the auxiliary task. It is actually a multi-task learning \cite{camana1993multitask} which has been demonstrated to be effective in recent DNN-based speech enhancement \cite{xu2015multi} and speech recognition \cite{huang2015rapid}.
%It also can be regarded as adding more constraints on the primary loss function.

\begin{figure}[t]
	\centering
	\centerline{\includegraphics[width=\columnwidth]{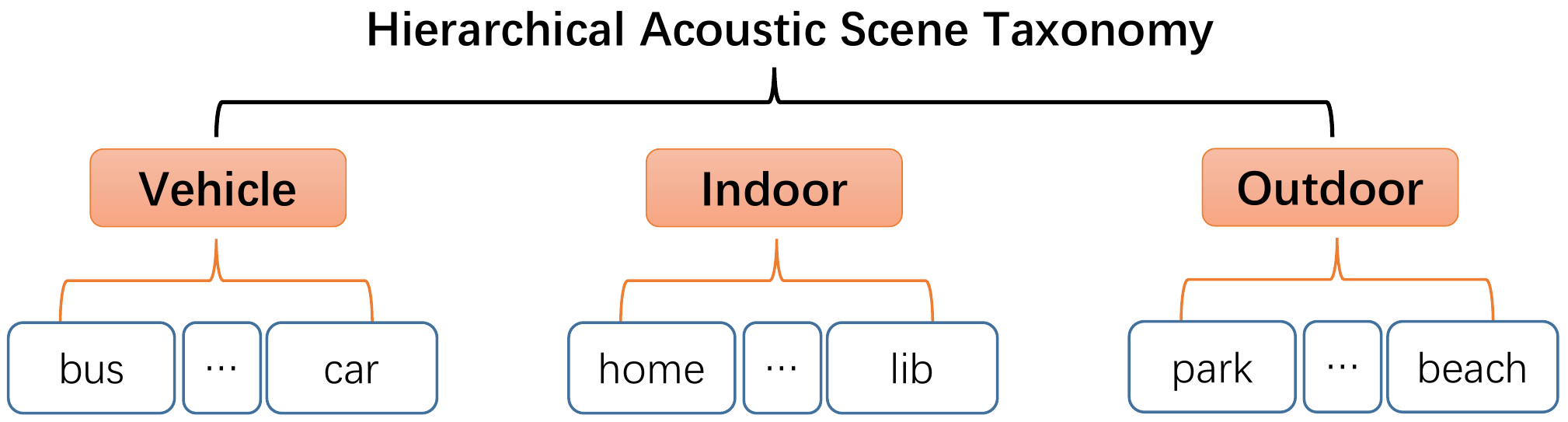}}
	\caption{Example of a hierarchical acoustic scene taxonomy.}
	\label{fig:taxonomy}
\end{figure}

The rest of the paper is organized as follows. In Section \ref{sec:dnn_baseline} we describe the DNN-based acoustic scene classification baseline system. Hierarchical pre-training and multi-level objective function are presented in Section \ref{sec:proposed_methods}. In Section \ref{sec:exp}, we present a series of experiments to assess the system performance. Finally we summarize our findings in Section \ref{sec:concluions}.

\section{DNN-based acoustic scene classification}
\label{sec:dnn_baseline}

DNN is a non-linear multi-layer model with powerful capability to extract robust feature related to a specific classification \cite{hinton2012deep} or regression \cite{xu2015regression} task. ASC is a typical classification problem where a specific scene label should be assigned to an audio segment.

\subsection{DNN baseline}
\label{ssec:baseline}
A basic DNN consists of a number of different layers stacked together in a deep architecture: an input layer, several hidden layers and an output layer. More precisely, when the goal is to classify an audio feature $ \textbf{x} $ among $ N $ acoustic scene classes, a DNN estimates the posteriors $ p_j$, $j\in\{1,...,J\}$, of each class given the input feature $ \textbf{x} $. The input $ \textbf{x} $ which is fed into DNN represents the contextual audio feature, such as 11 consecutive frames centered at the current frame. Such contextual information was shown to improve the prediction performance in DNN-based speech enhancement or speech recognition \cite{xu2015regression, hinton2012deep}. The activation functions used in each hidden unit of the hidden layers are non-linear sigmoid or Rectified Linear Units (ReLUs) \cite{dahl2013improving} function. The ReLU, which is adopted in this work, has several advantages over the sigmoid: faster computation and more efficient gradient propagation and it is defined below:
\begin{equation}
\label{eqn:Relu}
f(y)=\text{max}(0,y)
\end{equation}
where $y$ is the output of the hidden unit before activated by ReLU. The output is computed via the softmax nonlinearity to force the target label to have the maximum posterior while competing with other non-targets. The objective is to minimize the cross entropy between the predictions of DNN $ \textbf{p}=[p_1,...,p_J]^T $ and the target probabilities $ \textbf{d}=[d_1,...,d_J]^T $. The loss function is defined as follows:
\begin{equation}
\label{eqn:cross_entropy}
L=-\sum_{j=1}^{J}d_j\cdot\text{log}(p_j)
\end{equation}
The classical back-propagation (BP) algorithm \cite{hinton2012deep} can be used to update the weights and bias of DNN based on the calculated error.

\subsection{Dropout for the over-fitting problem}
\label{ssec:dropout}
Deep learning architectures have a natural tendency to over-fitting especially when there is a little training data. Dropout is a simple but effective way to alleviate this problem \cite{dahl2013improving}. In each training iteration, the feature value of every input unit and the activation of every hidden unit are randomly removed with a predefined probability (e.g., $ \rho $). These random perturbations effectively prevent the DNN from learning spurious dependencies. At the decoding stage, the DNN discounts all of the weights involved in the dropout training by $(1-\rho)$, regarded as a model averaging process \cite{hinton2012improving}.

For the acoustic scene classification task, the testing audio segment could be totally different from the used training audio segments due to the presence of background noise. Thus Dropout should be adopted to improve its robustness to generalize to variants of testing segments.

\subsection{Decision maker based on average confidence}
\label{ssec:decision_maker}
ASC aims to assign a single semantic label to an audio segment. Majority voting is often used to make a global decision across all of the single audio frames in this segment \cite{barchiesi2015acoustic}. Here we proposed to use a more precise decision making scheme:
\begin{equation}
\label{eqn:decision_maker}
\hat{c}=\max\limits_{j}\left(\frac{1}{T}\sum_{t=1}^{T}p_{t,j}\right) 
\end{equation}
where $T$ is the total number of frames belonging to the current testing audio segment, $\hat{c}$ denotes the predicted global scene label based on the average confidence across the whole frames, and $p_{t,j}$ represents the estimated DNN posterior at the $t$-th frame for class $j$.

\section{PROPOSED HIERARCHAL LEARNING FOR ASC}
\label{sec:proposed_methods}
In this section, we present two novel methods: hierarchical pre-training and multi-level objective function incorporating the scene taxonomy information for DNN-based ASC.

\subsection{Hierarchical pre-training}
\label{ssec:hier_pretr}

\begin{figure}[t]
	\centering
	\centerline{\includegraphics[width=\columnwidth]{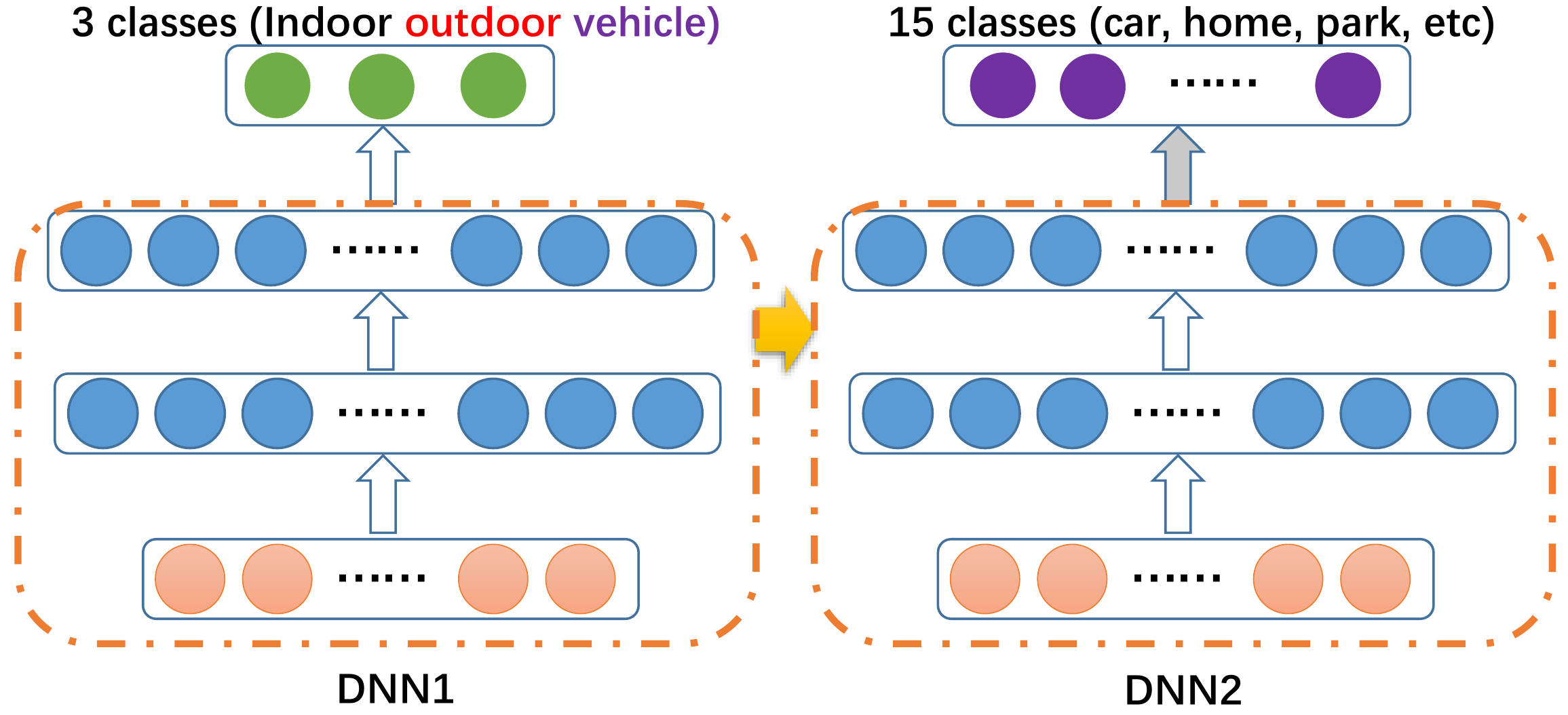}}
	\caption{Proposed hierarchical pre-training.}
	\label{fig:hier_pretr}
\end{figure}

Pre-training is crucial to avoid the algorithm getting stuck in a local optimum for training a deep model especially when the training data is not sufficient. The two most notable pre-training methods are the RBMs \cite{hinton2006reducing} based and stacked auto-encoders \cite{vincent2010stacked} based greedy layer-wise algorithms. They are both unsupervised while the proposed hierarchical pre-training is supervised. In the acoustic scene taxonomy research \cite{salamon2014dataset}, the acoustic scenes are naturally categorized into hierarchical classes. Fig. \ref{fig:hier_pretr} shows how the proposed DNN-based method incorporates the hierarchical taxonomy information. The hierarchical pre-training consists of two steps. Firstly, the DNN1 was trained to predict the three high-level acoustic scene classes, namely indoor, outdoor and vehicle. DNN2 was then trained to estimate the posterior of the 15 target low-level acoustic scene classes with the initialized weights from DNN1. Note that the classification layer of DNN2 was initialized with random weights because this top layer is different from DNN1. It is easier for DNN to learn the three coarsely classified high-level classes than the 15 target classes. However, the DNN2 can be better fine-tuned based on DNN1. It follows the common sense of human learning process: easiest things first, hardiest second. The experience of learning easier things could benefit the learning for harder things.

\subsection{Multi-level objective function}
\label{ssec:multi_obj}
Multi-task learning \cite{camana1993multitask} is successfully adopted in DNN-based speech enhancement \cite{xu2015multi} and DNN-based speech recognition \cite{huang2015rapid}. The auxiliary target was demonstrated beneficial for the primary target. Inspired by this, a multi-level objective function is proposed to incorporate the hierarchical acoustic scene taxonomy information into the integrated objective function.
\begin{figure}[t]
	\centering
	\centerline{\includegraphics[width=\columnwidth]{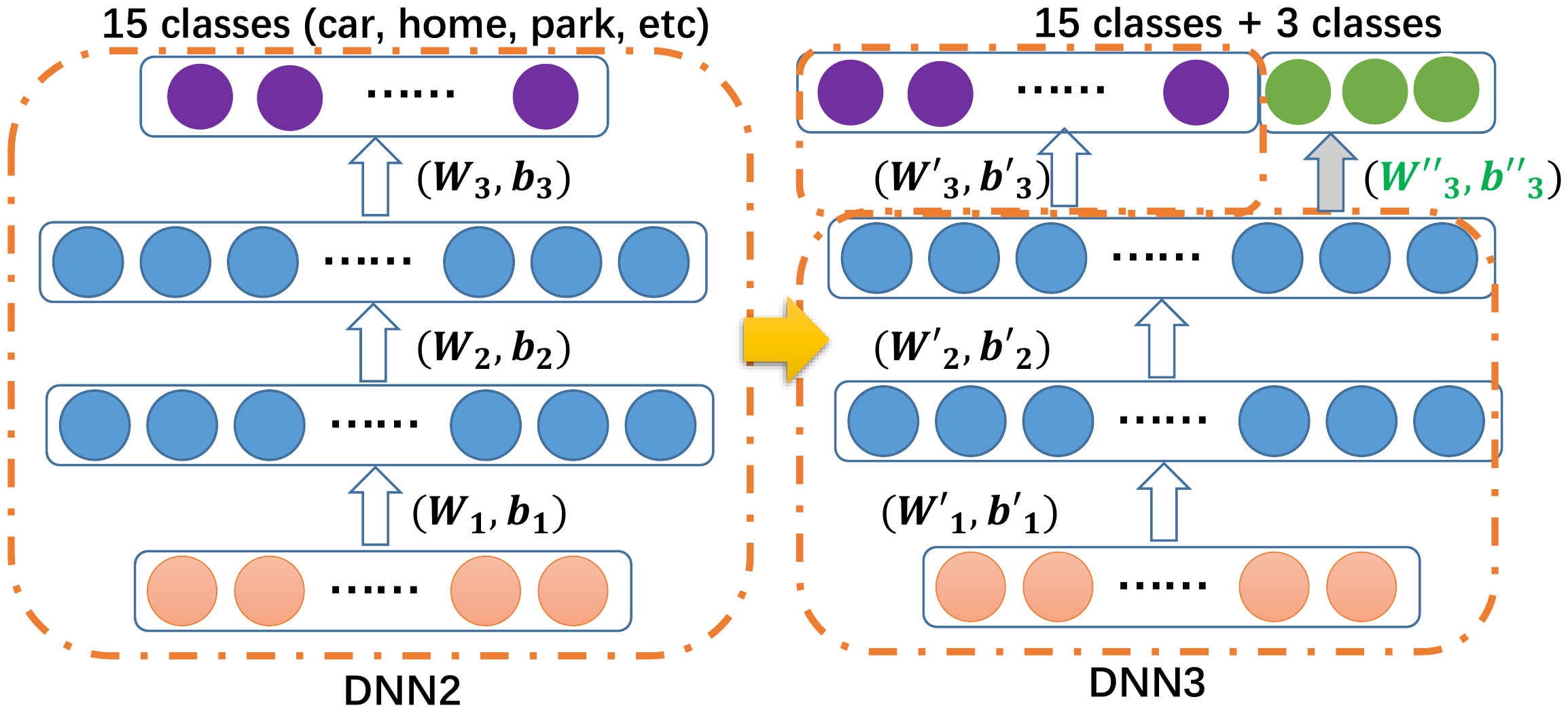}}
	\caption{Proposed multi-level objective function based on the well trained DNN2. $ \textbf{W} $ and $\textbf{b}$ denote the weights and bias, respectively.}
	\label{fig:multi_obj}
\end{figure}

Fig. \ref{fig:multi_obj} shows the proposed multi-level objective function based on the well trained DNN2. The main difference between DNN2 and DNN3 is that an additional softmax layer is designed to describe the three high-level classes (indoor, outdoor and vehicle). $ \textbf{W} $ and $\textbf{b}$ denote the weights and bias, respectively. $ \textbf{W}^\prime $ and $\textbf{b}^\prime$ of DNN3 were both initialized by $ \textbf{W} $ and $\textbf{b}$ of DNN2. The additional softmax layer $(\textbf{W}^{\prime\prime},\textbf{b}^{\prime\prime})$ was randomly initialized. With this modification, the cross entropy based loss function should be changed to contain two parts as follows:
\begin{equation}
\label{eqn:cross_entropy_multi}
L_{1:N}=-\alpha\sum_{t=1}^{N}\sum_{j=1}^{J}d_{t,j}\text{log}(p_{t,j})-(1-\alpha)\sum_{t=1}^{N}\sum_{k=1}^{K}d_{t,k}\text{log}(p_{t,k})
\end{equation}
where $N$ is the mini-batch size, $d_{t,j}$ denotes the target probability at the $t$-th frame for the $j$-th low-level scene class, $d_{t,k}$ denotes the DNN predicted posterior at the $t$-th frame for the $k$-th high-level scene class. $\alpha$ is the weighting factor to tune the error contribution from the above two parts. $J$ and $K$ represent the 15 low-level classes and the three high-level classes, respectively.

Hence, the proposed multi-level objective function is another idea to utilize the hierarchical scene taxonomy information besides the proposed pre-training in Sec. \ref{ssec:hier_pretr}.

\section{EXPERIMENTAL SETUP AND RESULTS}
\label{sec:exp}
The proposed methods were evaluated on the Task1 of DCASE 2016 challenge.
%Task1 is about acoustic scene classification aiming to classify a test recording into one of the predefined classes that characterizes the environment where it was recorded. The dataset consists of recordings from various acoustic scenes, all having distinct recording locations. For each recording location, 3-5 minutes long audio recording was captured. The original recordings were then split into 30-second segments for the challenge. 
There are 15 acoustic scenes for this task \footnote[1]{C1: Lakeside beach (outdoor); C2: Bus, traveling by bus in the city (vehicle); C3: Cafe / Restaurant, small cafe/restaurant (indoor); C4: Car, driving or traveling as a passenger (vehicle); C5: City center (outdoor); C6: Forest path (outdoor); C7: Grocery store, medium size grocery store (indoor); C8: Home (indoor);  C9: Library (indoor); C10: Metro station (indoor); C11: Office, multiple persons, typical work day (indoor); C12: Urban park (outdoor); C13: Residential area (outdoor); C14: Train (traveling, vehicle); C15: Tram (traveling, vehicle).}. Three high-level scene classes are also indicated.
For all of the acoustic scenes, each of the recordings was captured in a different location: different streets, different parks and different homes. Recordings were made using a Soundman OKM II Klassik/studio A3, electret binaural microphone and a Roland Edirol R-09 wave recorder using 44.1 kHz sampling rate and 24 bit resolution. The recordings are down-sampled into 16 kHz in this paper. The microphones are specifically made to look like headphones, being worn in the ears. As an effect of this, the recorded audio is very similar to the sound that reaches the human auditory system of the person wearing the equipment.

The dataset consists of two subsets: a development dataset and an evaluation dataset. In this paper, only the development dataset is used for evaluation because the labels of the evaluation dataset have not been released. The development dataset contains 1170 segments in total with 30 seconds length for each. A cross-validation setup with four folds is provided for the development dataset. The scoring of acoustic scene classification will be based on classification accuracy. Each segment is considered as an independent test sample. Confusion matrix among various acoustic scene classes would also be presented.

The official baseline system is based on the MFCC acoustic features and GMM classifier. The system learns one acoustic model per acoustic scene class, and performs the classification with maximum likelihood classification scheme. The length of each frame is 40 ms with 50\% hop size. The acoustic features include 20-dimension \text{MFCC} static coefficients (0th coefficient included), delta coefficients and acceleration coefficients.

For the DNN method, 11 frames of Mel-filter bank features with 40 channels were used as the input. Two hidden layers with 500 \text{ReLU} hidden units for each layer were adopted for DNN. The learning rate was 0.005. The momentum was set to 0.9. Weight cost was not used. The dropout value for the input layer was 0.1 while 0.3 for hidden layers. $\alpha$ in Eq. \ref{eqn:cross_entropy_multi} was 0.6. NVIDIA-Tesla-M2090 GPU was used to train the DNN models. The output unit number for DNN1, DNN2 and DNN3 were 3, 15 and 18, respectively.

\subsection{Evaluations for the proposed methods}
\label{ssec:eval_hier_learning}
As shown in Fig. \ref{fig:hier_pretr}, DNN1 should be trained as the pre-trained model for DNN2. Table \ref{tab:fr_acc_DNN1} gives the frame-wise accuracy (\%) for the three high-level scene classes on four cross-validation (CV) folds using DNN1. All of the related CV audio segments were excluded from the training sampels. An average of frame-level 90\% accuracy can be obtained for the classification of three high-level acoustic scene classes, namely indoor, outdoor and vehicle. Therefore, DNN can easily deal with this learning. It would offer a good starting optimization point for the fine-tuning of DNN2 with the initialized weights from DNN1.
\begin{table}[h]
	\begin{center}

		\begin{tabular}{|c|c|c|c|c|c|c|}
			\hline
			System & Fold 1 & Fold 2 & Fold 3 & Fold 4 & Average\\
			\hline
			DNN1 & 93\% & 90\% & 89\% & 91\% & 90\%\\
			\hline
		\end{tabular}
	\end{center}		
	\caption{Frame-wise accuracy (\%) for three high-level scene classes on different cross-validation (CV) folds using DNN1. All of the related CV audio segments were excluded from the training samples.}
	\label{tab:fr_acc_DNN1}
\end{table}
Then the DNN2 was trained to predict the 15 target acoustic scene classes based on the well trained DNN1. Table \ref{tab:all_DNNs} presented the overall comparison of acoustic scene accuracy (\%) on different CV folds among the DCASE2016 official GMM baseline, the DNN baseline improved by dropout, the DNN2 with the hierarchical pre-training based on DNN baseline, and the DNN3 optimized by the proposed multi-level objective function based on DNN2. All of the related CV audio segments were excluded from the training samples. The DNN baseline improved by dropout outperformed the provided GMM-MFCC baseline at all folds. The acoustic scene accuracy was increased from 71.28\% to 75.19\% on average. It also should be noted that the DNN is just slightly better than GMM on Fold 2 where the performance is the lowest. However, with the proposed hierarchical pre-training, its accuracy was significantly improved from 67.24\% to 71.72\% on Fold 2. Therefore, it demonstrates that the proposed hierarchical pre-training is important in challenging scene classification situations. DNN2 obtains an 8\% relative improvement compared with the DNN baseline from 75.19\% to 77.17\%. 
\begin{table*}
	\begin{center}
		\begin{tabular}{|c|c|c|c|c|c|c|}
			\hline
			Systems & Fold 1 ACC (\%) & Fold 2 ACC (\%) & Fold 3 ACC (\%) & Fold 4 ACC (\%) & Average ACC (\%)\\
			\hline
			GMM-baseline & 72.50 & 66.80 & 70.10 & 75.70 & 71.28\\
			\hline
			DNN-baseline (+dropout) & 79.62 & 67.24 & 75.84 & 78.08 & 75.19\\
			\hline
			DNN2 (+hierarchical pre-training) & 80.69 & 71.72 & 77.52 & 78.77 & 77.17\\
			\hline
			DNN3 (++multi-level objective func) & \textbf{81.38} & \textbf{72.41} & \textbf{77.85} & \textbf{79.79} & \textbf{77.86}\\
			\hline															
		\end{tabular}
	\end{center}
	\caption{The overall comparison of acoustic scene accuracy (\%) on different cross-validation (CV) folds among the DCASE2016 official GMM baseline, the DNN baseline improved by dropout, the DNN2 with the hierarchical pre-training based on DNN baseline, and the DNN3 optimized by the proposed multi-level objective function based on DNN2. All of the related CV audio segments were excluded from the training samples.}
	\label{tab:all_DNNs}
\end{table*}

\begin{figure}[t]
	\centering
	\centerline{\includegraphics[width=\columnwidth]{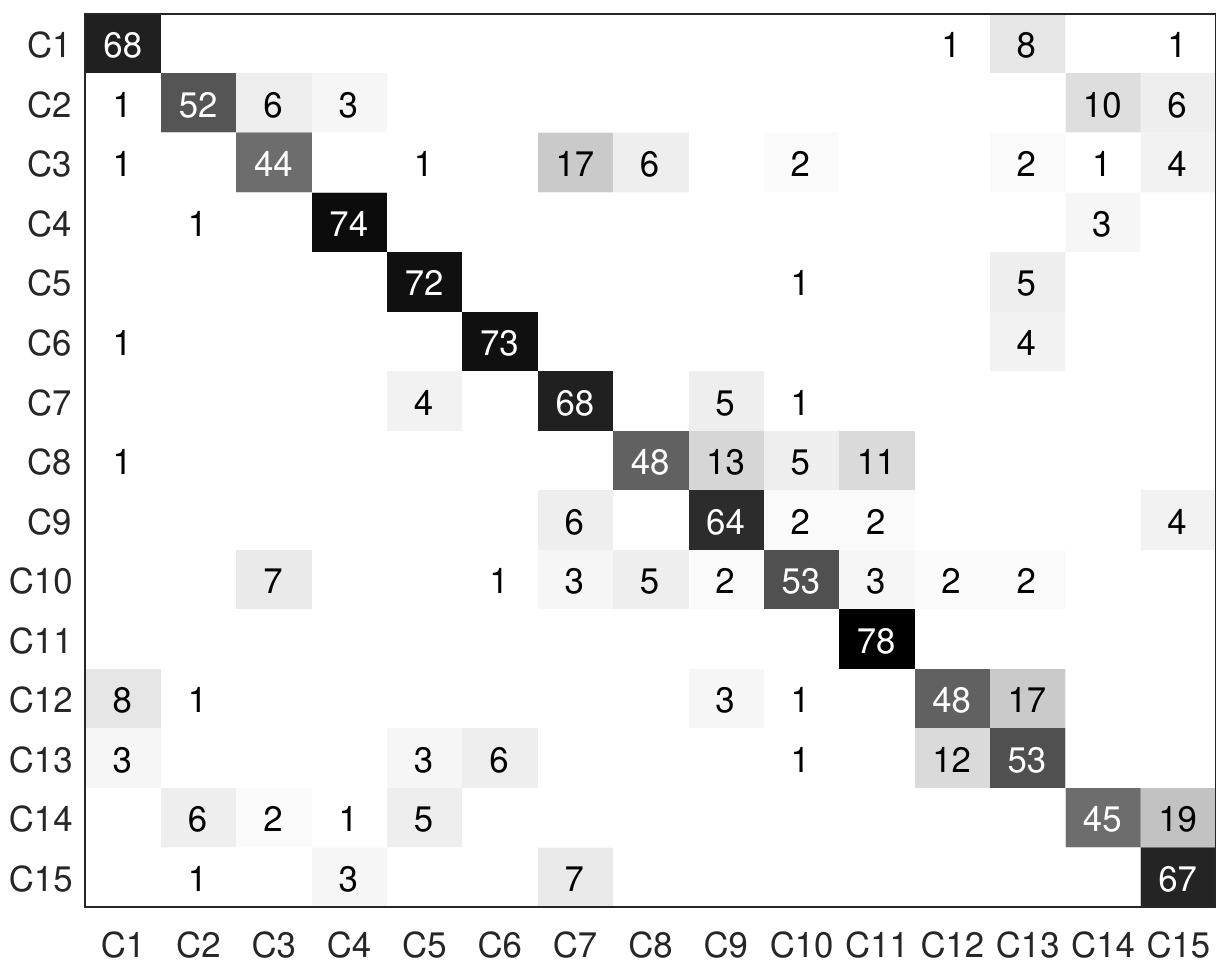}}
	\caption{The confusion matrix among 15 acoustic scene classes by comparing the DNN predicted class with the target class on all of the four folds. $Cx, x\in\{1,...,15\}$ represent the indices of the 15 classes which were defined in footnote 1.}
	\label{fig:conf_matrix}
\end{figure}

The DNN3 optimized by the proposed multi-level objective function gives further improvement. The final average acoustic scene accuracy was increased to 77.86\%. It indicates that the additional constraint imposed in Eq. \ref{eqn:cross_entropy_multi} can benefit the primary target. Finally, the proposed DNN system offers 22.9\% and 10.8\% relative improvements compared with the GMM-MFCC baseline and the DNN baseline, respectively. Note that the DNN baseline is a strong system since it is optimized by dropout training.

\subsection{Further discussions}
\label{ssec:discussion}
Fig. \ref{fig:conf_matrix} presents the confusion matrix among 15 acoustic scene classes by comparing the DNN predicted class with the target class on all of the four folds. $Cx, x\in\{1,...,15\}$ represents the indices of the 15 classes which were defined in footnote 1. Observed from this confusion matrix, one phenomenon is that \textit{park} ($C12$) easily gets confused by \textit{residential area} ($C13$), and vice versa. It could be explained that similar acoustic events happened in both acoustic environments, like the \textit{bird singing} and \textit{car passing-by}. Another interesting case is that \textit{grocery store} ($C7$) tends to be mis-recognized as \textit{restaurant} ($C3$) due to the common human speech events. This might suggest that the presence of common human speech needs to be reduced in the audio segments before the acoustic scene classification is conducted. \textit{Tram} ($C15$) also has the tendency to be incorrectly identified as \textit{Train} ($C14$).
%In this acoustic scene classification task, the characteristics of the test audio segment can be very different from the used training audio segments because of the randomly happening acoustic events. The adopted Dropout method can alleviate this problem. However, more robust feature learning methods should be developed to extract the specific acoustic characteristics of the certain acoustic environments.
\begin{table}[h]
	\begin{center}
		\begin{tabular}{|c|c|c|}
			\hline
			System & Proposed method & DCASE2016 Baseline \\
			\hline
			acc & 80.5\% & 77.2\% \\
			\hline
		\end{tabular}
	\end{center}
	\caption{Accuracy (\%) for the final evaluation set.}
	\label{tab:acc_evaluation}
\end{table}
Table \ref{tab:acc_evaluation} presents the final accuracy for the evaluation set. The final DNN model was trained with the whole 1170 segments of the development set. Note that the proposed method can only get 73.3\% accuracy if the DNN was trained on Fold1 only (as given on DCASE2016 Task1 website).

\section{CONCLUSIONS}
\label{sec:concluions}
%The hierarchical structure of the acoustic scene taxonomy is natural shown in Fig. \ref{fig:taxonomy}.
In this paper, we have studied how to incorporate the taxonomy information into deep learning framework, and developed two DNN-based hierarchical learning methods for the acoustic scene classification task. The first novel method, called hierarchical pre-training which is a supervised learning process, can help the second DNN to get a better initialized weights based on the learning experience from the three high-level coarsely classified classes. It can achieve an 8\% relative improvement compared with the DNN baseline improved by Dropout. The second proposed approach was the multi-level objective function which was inspired by the multi-task learning. It can help improve the prediction accuracy of the primary 15 target low-level classes by adding additional estimation of the three high-level classes in the DNN output, which was also regarded as imposing more constraint on the cross-entropy loss function. This idea can further improve the scene classification performance. Finally, the proposed DNN system has obtained 22.9\% and 10.8\% relative improvements over the GMM-MFCC baseline and the well trained DNN baseline, respectively.

%In our future work, the hierarchical learning will also be investigated in more complicated network structure, such as the CNN and the long-short term memory (LSTM) model  based framework, for the acoustic scene classification task.

\newpage
% -------------------------------------------------------------------------
% Either list references using the bibliography style file IEEEtran.bst
\bibliographystyle{IEEEtran}
\bibliography{refs}

\begin{thebibliography}{10}
\providecommand{\url}[1]{#1}
\csname url@samestyle\endcsname
\providecommand{\newblock}{\relax}
\providecommand{\bibinfo}[2]{#2}
\providecommand{\BIBentrySTDinterwordspacing}{\spaceskip=0pt\relax}
\providecommand{\BIBentryALTinterwordstretchfactor}{4}
\providecommand{\BIBentryALTinterwordspacing}{\spaceskip=\fontdimen2\font plus
\BIBentryALTinterwordstretchfactor\fontdimen3\font minus
  \fontdimen4\font\relax}
\providecommand{\BIBforeignlanguage}[2]{{%
\expandafter\ifx\csname l@#1\endcsname\relax
\typeout{** WARNING: IEEEtran.bst: No hyphenation pattern has been}%
\typeout{** loaded for the language `#1'. Using the pattern for}%
\typeout{** the default language instead.}%
\else
\language=\csname l@#1\endcsname
\fi
#2}}
\providecommand{\BIBdecl}{\relax}
\BIBdecl

\bibitem{barchiesi2015acoustic}
D.~Barchiesi, D.~Giannoulis, D.~Stowell, and M.~D. Plumbley, ``Acoustic scene
  classification: Classifying environments from the sounds they produce,''
  \emph{IEEE Signal Processing Magazine}, vol.~32, no.~3, pp. 16--34, 2015.

\bibitem{mesaros2010acoustic}
A.~Mesaros, T.~Heittola, A.~Eronen, and T.~Virtanen, ``Acoustic event detection
  in real life recordings,'' in \emph{IEEE 18th European Signal Processing
  Conference}, 2010, pp. 1267--1271.

\bibitem{zhuang2010real}
X.~Zhuang, X.~Zhou, M.~A. Hasegawa-Johnson, and T.~S. Huang, ``Real-world
  acoustic event detection,'' \emph{Pattern Recognition Letters}, vol.~31,
  no.~12, pp. 1543--1551, 2010.

\bibitem{sawhney1997situational}
N.~Sawhney and P.~Maes, ``Situational awareness from environmental sounds,''
  \emph{{URL}: http://web. media. mit. edu/\~{} nitin/papers/Env\_Snds/EnvSnds.
  html}, 1997.

\bibitem{giannoulis2013ieee}
D.~Giannoulis, E.~Benetos, D.~Stowell, M.~Rossignol, M.~Lagrange, and
  M.~Plumbley, ``{IEEE AASP} challenge: Detection and classification of
  acoustic scenes and events,'' Queen Mary University of London, Tech. Rep.,
  2013.

\bibitem{giannoulis2013detection}
D.~Giannoulis, E.~Benetos, D.~Stowell, M.~Rossignol, M.~Lagrange, and M.~D.
  Plumbley, ``Detection and classification of acoustic scenes and events: An
  {IEEE AASP} challenge,'' in \emph{2013 IEEE Workshop on Applications of
  Signal Processing to Audio and Acoustics (WASPAA)}, 2013, pp. 1--4.

\bibitem{stowell2015detection}
D.~Stowell, D.~Giannoulis, E.~Benetos, M.~Lagrange, and M.~D. Plumbley,
  ``Detection and classification of acoustic scenes and events,'' \emph{IEEE
  Transactions on Multimedia}, vol.~17, no.~10, pp. 1733--1746, 2015.

\bibitem{geiger2013large}
J.~T. Geiger, B.~Schuller, and G.~Rigoll, ``Large-scale audio feature
  extraction and svm for acoustic scene classification,'' in \emph{IEEE
  Workshop on Applications of Signal Processing to Audio and Acoustics
  (WASPAA)}, 2013, pp. 1--4.

\bibitem{lee2013acoustic}
K.~Lee, Z.~Hyung, and J.~Nam, ``Acoustic scene classification using sparse
  feature learning and event-based pooling,'' in \emph{IEEE Workshop on
  Applications of Signal Processing to Audio and Acoustics (WASPAA)}, 2013, pp.
  1--4.

\bibitem{bisot2016acoustic}
V.~Bisot, R.~Serizel, S.~Essid \emph{et~al.}, ``Acoustic scene classification
  with matrix factorization for unsupervised feature learning,'' in \emph{2016
  IEEE International Conference on Acoustics, Speech and Signal Processing
  (ICASSP)}, 2016, pp. 6445--6449.

\bibitem{xu2014experimental}
Y.~Xu, J.~Du, L.-R. Dai, and C.-H. Lee, ``An \text{experimental study} on
  speech enhancement based on deep neural networks,'' \emph{IEEE Signal
  Processing Letters}, vol.~21, no.~1, pp. 65--68, 2014.

\bibitem{xu2015regression}
------, ``A regression approach to speech enhancement based on deep neural
  networks,'' \emph{IEEE/ACM Transactions on Audio, Speech, and Language
  Processing}, vol.~23, no.~1, pp. 7--19, 2015.

\bibitem{hinton2012deep}
G.~Hinton, L.~Deng, D.~Yu, G.~E. Dahl, A.-r. Mohamed, N.~Jaitly, A.~Senior,
  V.~Vanhoucke, P.~Nguyen, T.~N. Sainath \emph{et~al.}, ``Deep neural networks
  for acoustic modeling in speech recognition: The shared views of four
  research groups,'' \emph{IEEE Signal Processing Magazine}, vol.~29, no.~6,
  pp. 82--97, 2012.

\bibitem{krizhevsky2012imagenet}
A.~Krizhevsky, I.~Sutskever, and G.~E. Hinton, ``Imagenet classification with
  deep convolutional neural networks,'' in \emph{Advances in {N}eural
  {I}nformation {P}rocessing {S}ystems}, 2012, pp. 1097--1105.

\bibitem{hinton2006reducing}
G.~E. Hinton and R.~R. Salakhutdinov, ``Reducing the dimensionality of data
  with neural networks,'' \emph{Science}, vol. 313, no. 5786, pp. 504--507,
  2006.

\bibitem{petetin2015deep}
Y.~Petetin, C.~Laroche, and A.~Mayoue, ``Deep neural networks for audio scene
  recognition,'' in \emph{{IEEE} 23rd European Signal Processing Conference
  (EUSIPCO)}, 2015, pp. 125--129.

\bibitem{piczak2015environmental}
K.~J. Piczak, ``Environmental sound classification with convolutional neural
  networks,'' in \emph{2015 IEEE 25th International Workshop on Machine
  Learning for Signal Processing (MLSP)}, 2015, pp. 1--6.

\bibitem{ravanelli2014audio}
M.~Ravanelli, B.~Elizalde, K.~Ni, and G.~Friedland, ``Audio concept
  classification with hierarchical deep neural networks,'' in \emph{2014
  Proceedings of the 22nd European Signal Processing Conference ({EUSIPCO})},
  2014, pp. 606--610.

\bibitem{niessen2010categories}
M.~Niessen, C.~Cance, and D.~Dubois, ``Categories for soundscape: {T}oward a
  hybrid classification,'' in \emph{Inter-Noise and Noise-Con Congress and
  Conference Proceedings}, vol. 2010, no.~5, 2010, pp. 5816--5829.

\bibitem{salamon2014dataset}
J.~Salamon, C.~Jacoby, and J.~P. Bello, ``A dataset and taxonomy for urban
  sound research,'' in \emph{Proceedings of the ACM International Conference on
  Multimedia}, 2014, pp. 1041--1044.

\bibitem{lee2011learning}
Y.~J. Lee and K.~Grauman, ``Learning the easy things first: Self-paced visual
  category discovery,'' in \emph{2011 IEEE Conference on Computer Vision and
  Pattern Recognition (CVPR)}, 2011, pp. 1721--1728.

\bibitem{camana1993multitask}
R.~Camana, ``Multitask learning: A knowledge-based source of inductive bias,''
  in \emph{Proceedings of the Tenth International Conference on Machine
  Learning}, 1993, pp. 41--48.

\bibitem{xu2015multi}
Y.~Xu, J.~Du, Z.~Huang, L.-R. Dai, and C.-H. Lee, ``Multi-objective learning
  and mask-based post-processing for deep neural network based speech
  enhancement,'' in \emph{Sixteenth Annual Conference of the International
  Speech Communication Association {INTERSPEECH}}, 2015.

\bibitem{huang2015rapid}
Z.~Huang, J.~Li, S.~M. Siniscalchi, I.-F. Chen, J.~Wu, and C.-H. Lee, ``Rapid
  adaptation for deep neural networks through multi-task learning,'' in
  \emph{Sixteenth Annual Conference of the International Speech Communication
  Association}, 2015.

\bibitem{dahl2013improving}
G.~E. Dahl, T.~N. Sainath, and G.~E. Hinton, ``Improving deep neural networks
  for {LVCSR} using rectified linear units and dropout,'' in \emph{2013 IEEE
  International Conference on Acoustics, Speech and Signal Processing
  (ICASSP)}, 2013, pp. 8609--8613.

\bibitem{hinton2012improving}
G.~E. Hinton, N.~Srivastava, A.~Krizhevsky, I.~Sutskever, and R.~R.
  Salakhutdinov, ``Improving neural networks by preventing co-adaptation of
  feature detectors,'' \emph{arXiv preprint arXiv:1207.0580}, 2012.

\bibitem{vincent2010stacked}
P.~Vincent, H.~Larochelle, I.~Lajoie, Y.~Bengio, and P.-A. Manzagol, ``Stacked
  denoising autoencoders: Learning useful representations in a deep network
  with a local denoising criterion,'' \emph{The Journal of Machine Learning
  Research}, vol.~11, pp. 3371--3408, 2010.

\end{thebibliography}
%
% or list them by yourself
% \begin{thebibliography}{9}
% 
% \bibitem{dcase2016web}
%   \url{http://www.cs.tut.fi/sgn/arg/dcase2016/}.
%
% \bibitem{IEEEPDFSpec}
%   {PDF} specification for {IEEE} {X}plore$^{\textregistered}$,
%   \url{http://www.ieee.org/portal/cms_docs/pubs/confstandards/pdfs/IEEE-PDF-SpecV401.pdf}.
%
% \bibitem{PDFOpenSourceTools}
%   Creating high resolution {PDF} files for book production with 
%   open source tools, 
%   \url{http://www.grassbook.org/neteler/highres_pdf.html}.
%
% \bibitem{eWilliams1999}
% E. Williams, \emph{Fourier Acoustics: Sound Radiation and Nearfield Acoustic
%   Holography}. London, UK: Academic Press, 1999.
% 
% \bibitem{ieeecopyright}
%   \url{http://www.ieee.org/web/publications/rights/copyrightmain.html}.
%
% \bibitem{cJones2003}
% C. Jones, A. Smith, and E. Roberts, ``A sample paper in conference
%   proceedings,'' in \emph{Proc. IEEE ICASSP}, vol. II, 2003, pp. 803--806.
% 
% \bibitem{aSmith2000}
% A. Smith, C. Jones, and E. Roberts, ``A sample paper in journals,'' 
%   \emph{IEEE Trans. Signal Process.}, vol. 62, pp. 291--294, Jan. 2000.
% 
% \end{thebibliography}

\end{sloppy}
\end{document}